\documentclass[12pt]{article}

\usepackage{scicite}
\usepackage{times}
\usepackage{url}
\usepackage[version=4]{mhchem}
\usepackage{siunitx}
\usepackage{graphicx}
\usepackage{todonotes}
\usepackage{titling}
\usepackage{xpatch}
\makeatletter
\newenvironment{mytitlepage}%
  {\begin{titlepage}\def\@thanks{}}%
  {\end{titlepage}}
\xpatchcmd\titlepage{\setcounter{page}\@ne}{}{}{}
\xpatchcmd\endtitlepage{\setcounter{page}\@ne}{}{}{}
\makeatother
\newcommand{\beginsupplement}{%
        \setcounter{table}{0}
        \renewcommand{\thetable}{S\arabic{table}}%
        \setcounter{figure}{0}
        \renewcommand{\thefigure}{S\arabic{figure}}%
     }

\newenvironment{sciabstract}{%
\begin{quote} \bf}
{\end{quote}}

\title{Including climate system feedbacks in calculations of the social cost of methane}

\author
{Kristina R. Colbert,$^{1\ast}$ Frank C. Errickson,$^{2}$ David Anthoff$^{2}$, \\ Chris E. Forest$^{1}$
\\
\normalsize{$^{1}$Department of Meteorology and Atmospheric Science, The Pennsylvania State University,}\\ \normalsize{State College, PA}\\
\normalsize{$^{2}$Energy and Resources Group, University of California-Berkeley, Berkeley, CA,}\\
\\
\normalsize{$^\ast$Correspondence to: cef13@psu.edu.}
}

\date{}

\begin{document}
\begin{mytitlepage}


\baselineskip24pt


\maketitle



\begin{sciabstract}
Integrated assessment models (IAMs) are valuable tools that consider the interactions between socioeconomic systems and the climate system. Decision-makers and policy analysts employ IAMs to calculate the marginalized monetary cost of climate damages resulting from an incremental emission of a greenhouse gas. Used within the context of regulating anthropogenic methane emissions, this metric is called the social cost of methane (SC-\ce{CH4}). Because several key IAMs used for social cost estimation contain a simplified model structure that prevents the endogenous modeling of non-\ce{CO2} greenhouse gases, very few estimates of the SC-\ce{CH4} exist. For this reason, IAMs should be updated to better represent methane cycle dynamics that are consistent with comprehensive Earth System Models. We include feedbacks of climate change on the methane cycle to estimate the SC-\ce{CH4}. Our expected value for the SC-\ce{CH4} is \$1163/t-\ce{CH4} under a constant 3.0\% discount rate. This represents a 44\% increase relative to a mean estimate without feedbacks on the methane cycle.
\end{sciabstract}


\paragraph* 
\noindent \textit{Including methane feedbacks on biogeochemical processes and atmospheric chemistry increases the social cost of methane.\\}



The U.S. government Interagency Working Group (IWG) released updated values of the social cost of \ce{CO2} (SCC) in 2017 \cite{IWG2017}, but it did not provide estimates for the social cost of other greenhouse gases that would reflect the best available science and treatment of uncertainty. Although several social cost of methane (SC-\ce{CH4}) estimates exist in the literature \cite{Shindell2017, Marten2015, Waldhoff2014, Marten2012, Hope2006}, we consider four deficiencies in their model structures and assumptions. First, previous studies have converted methane into a \ce{CO2}-equivalent values using the 100-year global warming potential (GWP\textsubscript{100}) \cite{Lashof1990} and then multiply this result by the SCC to derive the SC-\ce{CH4} \cite{Kumari2019}. This method fails to capture the temporal nature of short-lived greenhouse gases, like methane, on the global temperature response. Second, weighting the GWP\textsubscript{100} over time fails to reflect how damages from climate change are likely to increase faster than the rise of the global mean temperature \cite{Bruhl1993}. Third, several studies of the SC-\ce{CH4} fail to account for important climate model parametric uncertainties in their analysis \cite{Marten2012, Hope2006}. Finally, others have not been calibrated to observations of the climate record and have not demonstrated that they can replicate key behaviors of comprehensive Earth System Models \cite{IWG2017, Harmsen2015}. 

Only a select few studies are consistent with the estimation method of the SCC. The work by Marten et al. \cite{Marten2015} is recognized by the U.S. Government for estimating the SC-\ce{CH4} using the same three IAMs, socioeconomic scenarios, time-horizon, and discount rates as the official U.S. SCC estimates \cite{IWG2017}. One shortcoming of their study is it assumes a constant methane lifetime box model for estimating methane concentrations. When compared to an analogous projection from the more comprehensive simple climate model, MAGICC \cite{Meinshausen2011b}, the methane concentration from the box model differs by 15\% in the first year following a pulse emission. The difference is magnified by the radiative forcing as the constant lifetime assumption poorly represents how concentrations of other gases and atmospheric temperature impact the methane lifetime. We argue that reduced-form methane cycle models used for the calculation of the social cost of methane should include the most significant mechanisms that impact methane's lifetime.

Complex Earth System Models with detailed atmospheric chemistry simulate the change in methane lifetime as variations in the methane sources and sinks directly impact how long methane stays in the atmosphere \cite{Prather1994}. Methane's primary atmospheric sink is the oxidation reaction with tropospheric hydroxyl radical (OH). As methane and other reactant species accumulate in the atmosphere, OH radicals are suppressed. As a result, the rate at which methane oxidizes slows down and methane's residence time increases. Since the doubling of preindustrial methane concentration, it is estimated that the increasing feedback has extended methane's lifetime from 5.9 to 9.2 years \cite{Dlugokencky2019,Holmes2018}. Projections indicate methane lifetime could increase by about 13\% this century, where the sensitivity is dominated by methane's feedback on its sink \cite{Holmes2013}.

A second important methane feedback relates to the temperature sensitivity og natural wetland systems. Wetlands are the largest source of natural methane emissions, releasing 150 to 225 Tg-\ce{CH4} each year or approximately a third of total global methane emissions \cite{Saunois2016}. The production of methane in these environments is predominately due to the degradation of organic matter by methanogen microbes in anaerobic (low-oxygen) conditions \cite{Ferry1999}. It is well known that temperature controls the rate of microbiological processes, and the production and oxidation of methane in wetland environments \cite{Kip2010,Frenzel2000}. With projected warming, rising surface air temperatures will pervade the top soil layers where the majority of methanotrophs reside, thus increasing the methane emitted from the wet soils. If natural methane emissions were to increase with the changing climate, the amplified emissions from these systems would further induce an increase in atmospheric concentration and lead to a positive climate feedback.

To update the SC-\ce{CH4} we include two important climate system feedbacks on the methane cycle that past estimates do not account for. We introduce the chemical feedback on methane lifetime and the wetland emission feedback into the methane component of a reduced-complexity climate model, FaIR v1.3 (Finite Amplitude Impulse Response) \cite{Smith2018}, which is a modified version of the Intergovernmental Panel on Climate Change impulse-response climate-carbon cycle model \cite{Rogelj2018} (see Methods). FaIR does not include climate feedbacks on the methane cycle. We use a Bayesian framework to calibrate our improved methane cycle model to past observations ad well as the feedback behavior simulated in a higher complexity model, the MIT Earth System Model (MESM) \cite{Sokolov2018}.

We replace the methane cycle component of the FUND integrated assessment model (IAM) (The Climate Framework for Uncertainty, Negotiation and Distribution) \cite{Anthoff2014} with our improved methane module. FUND adopts simplified representations of socioeconomic components, climate dynamics, and impact analysis. Compared to other IAMs used by the U.S. Government, it differs in its more detailed depiction of sectoral and regional economic impacts, as well as its endogenous greenhouse gas calculations. We use a Markov chain Monte Carlo (MCMC) algorithm to produce 10,000 samples from the joint posterior distribution of the uncertain methane cycle model parameters within FUND. By sampling the parameter space of the modified methane module, we account for a previously neglected source of climate model uncertainty in our SC-\ce{CH4} estimates. For each posterior parameter sample, we estimate the SC-\ce{CH4} using the constant consumption discount rates (2.5\%, 3.0\%, and 5.0\%) from official U.S. SC-\ce{CH4} estimates.

Under the  highest discount rate of 5.0\%, we estimate a mean SC-CH4 of \$408 (\$150-1076, 5-95\% range). In comparison, the lowest discount rate of 2.5\% increases the SC-CH4 to \$1492 (\$262-3331, 5-95\% range). Distributions for our SC-\ce{CH4} estimates at the 3.0\% constant discount rate are shown in Figure \ref{Compare_SCCH4_3per}. The expected value is \$1163 (\$253-2624, 5-95\% range) (2007 U.S. dollars per t-\ce{CH4}) and the distribution also contains a long upper-tail. However, when methane feedbacks are turned off, the SC-\ce{CH4} distribution is shifted towards lower values, with a shorter upper-tail and expected value of \$806. Our results indicate that adding methane feedbacks to the FUND climate model increases our mean SC-\ce{CH4} estimate by 44\%. Further, the 95\textsuperscript{th} percentile, chosen to represent potential higher-than-expected impacts of methane, is 45\% higher when feedbacks are included in the calculation of the SC-\ce{CH4}. The feedbacks create a more damaging effect because a small increase in methane emissions prolongs the methane atmospheric residence time and produces an additional radiative impact.

The marginal atmospheric methane concentration and discounted monetary damages (3\% discount rate) following a unit ton emission pulse of methane are displayed in Figure \ref{marginalresponses}. There are a few distinct changes in the methane concentration response after updating the methane cycle in FUND. The peak in atmospheric methane concentration is 14\% higher for the updated model because natural emissions from wetlands increase the overall total emissions. A second notable difference is the decay rate of the perturbed concentration anomaly. In the default model without any methane feedbacks, the peak concentration drops 50\% over 8 years. Switching the FUND model to include a dynamic methane lifetime dependent on concentration extends the lifetime following a unit increase in emission. The overall increased lifetime slows the decay so that half of the perturbed concentration anomaly takes nearly 13 years to be removed from the atmosphere.

Direct comparison with previous estimates is difficult, given the differences in IAM model choice and socioeconomic emission scenarios. To make a fair comparison with other studies, the emission scenario, time horizon, IAM, and choice in discount rate must be the same. The work referenced by the U.S. IWG, \cite{Marten2015}, contains the closest comparison. Our expected value of \$1163, falls within their SC-\ce{CH4} range of \$980--1,400 using FUND with a 3\% constant discount rate. Albeit, their study used four different emission scenarios that spanned emissions lower and greater than that used within our analysis. Because our estimates use the default emission scenario from FUND, it is difficult to directly compare to estimates using alternative emission scenarios. The choice in emission scenario would ultimately affect the total radiative forcing and temperature response of the model.

Our study utilizes FUND, a prominent IAM, for the calculation of the SC-\ce{CH4}. Including climate feedbacks on methane concentrations in our new SC-\ce{CH4} estimates advances the work of previous studies in a manner more consistent with the methodology of the SCC. As a potent greenhouse gas, methane will produce a stronger temperature response with climate system feedbacks included. That temperature response would then equate to greater climate damages related to agricultural productivity, human health, changes in energy system costs, and ecosystem destruction. Monetized climate damages could reach \$2624/t-\ce{CH4} under a high impact scenario (95\% confidence bound), where climate system feedbacks on the methane cycle lead to a 45\% increase in the SC-\ce{CH4}.






\section*{Acknowledgments}
\textbf{Funding:} This work was partially supported by the National Science Foundation through the Network for
Sustainable Climate Risk Management (SCRiM) under NSF cooperative agreement GEO-1240507. Any opinions, findings, and conclusions or recommendations expressed in this material are those of the authors and do not necessarily reflect the views of the National Science Foundation. \textbf{Author contributions:} 
K.R.C wrote the original draft and performed the formal analysis. F.C.E and D.A. coded and maintained the models. C.E.F. and D.A. provided project administration and supervision. All authors contributed to the conceptualization, methodology, and interpretation of results. \textbf{Competing interests:} Authors declare no competing interests. \textbf{Data and materials availability:} The model data will be available at the Penn State Data Commons (www.datacommons.psu.edu). Models are open sourced on GitHub (github.com/mimiframework).




\section*{List of Supplementary materials}
Methods\\
\indent Target Data Set for FaIR Calibration \\
\indent Updating the FaIR Natural Methane Emissions \\ 
\indent Updating the FaIR Methane Lifetime \\
\indent FaIR Calibration Process \\
\indent Uncertainty Estimation \\
\indent Social Cost of Methane Calculation \\
Supplementary Text \\
\indent Assessment of the Updated FaIR Methane Module \\
Supplementary Figures S1 to S5 \\
References (22--42) \\




\newpage
\noindent {\bf Fig. 1.} A comparison of the SC-\ce{CH4} distribution for the default version 3.9 of FUND (teal) and the modified version containing methane feedbacks (purple) under a 3.0\% constant discount rate assumption. Dashed lines indicate the mean SC-\ce{CH4} estimate for each distribution. 

\noindent {\bf Fig. 2.} Impulse response behavior and uncertainties following a one-ton methane pulse in 2020 using the FUND integrated assessment model. The solid lines indicate the mean response for 500 randomly selected members of the 10,000 model runs. A) Atmospheric methane concentration response for the default FUND version 3.9 (teal)  and modified FUND with methane feedbacks (purple). B) Predicted discounted climate damage using a 3.0\% constant consumption discount rate.



\newpage
\begin{figure}[htp!]
    \centering
    \includegraphics[width=\textwidth]{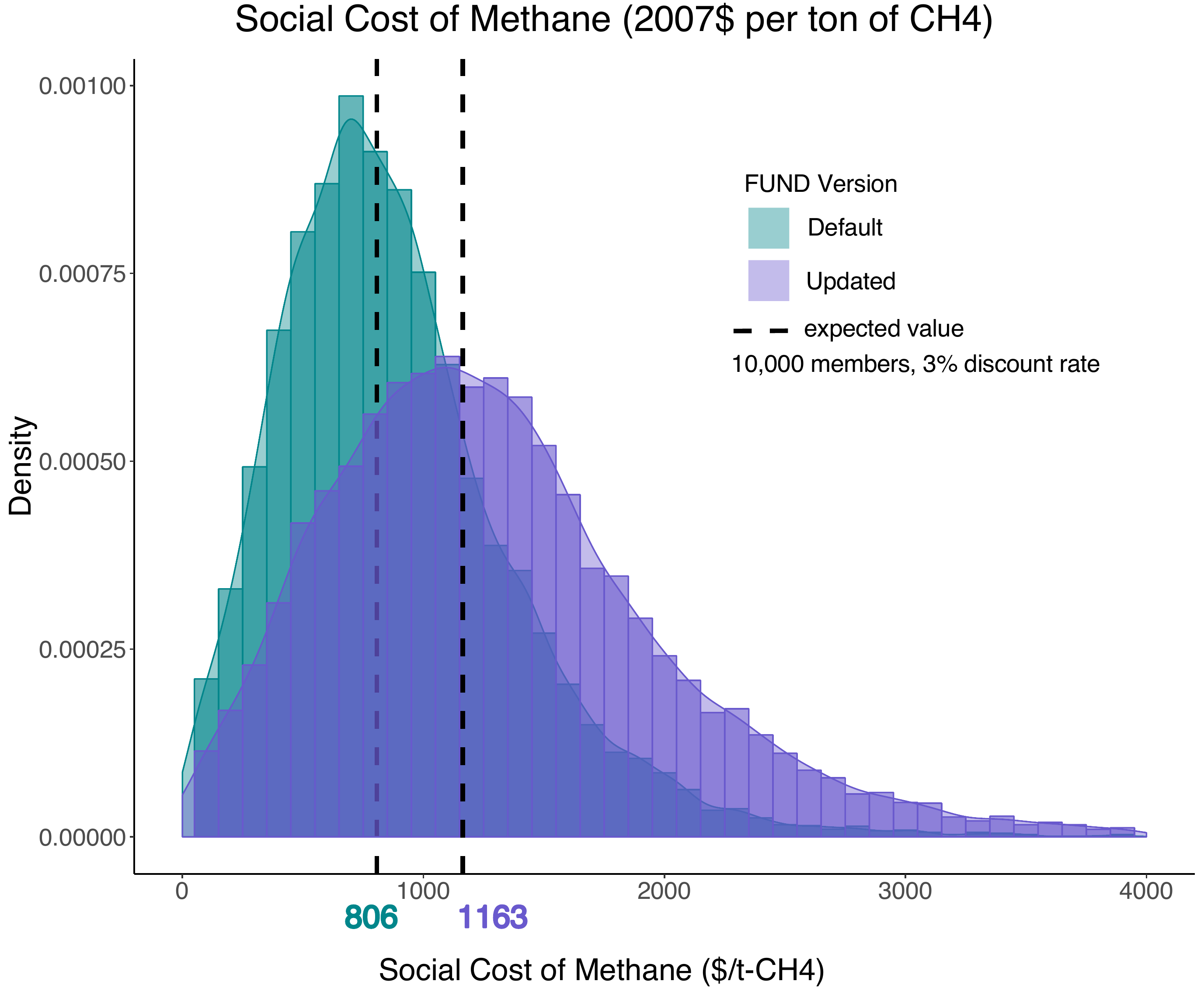}
    \caption{A comparison of the SC-\ce{CH4} distribution for the default version 3.9 of FUND (teal) and the modified version containing methane feedbacks (purple) under a 3.0\% constant discount rate assumption. Dashed lines indicate the mean SC-\ce{CH4} estimate for each distribution.}
    \label{Compare_SCCH4_3per}
\end{figure}

\newpage
\begin{figure}[htp!]
    \centering
    \includegraphics[width=0.6\textwidth]{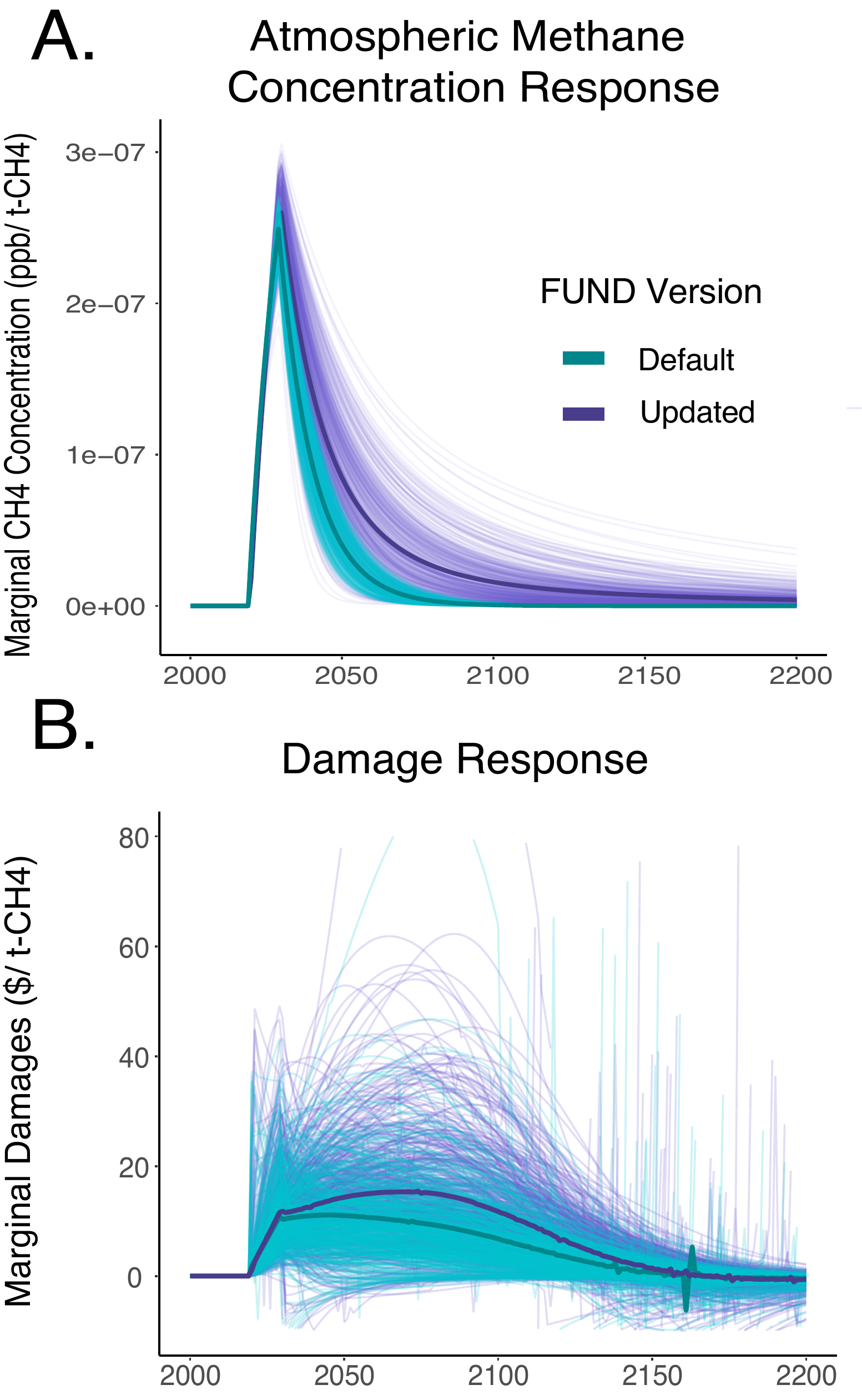}
    \caption{Impulse response behavior and uncertainties following a one-ton methane pulse in 2020 using the FUND integrated assessment model. The solid lines indicate the mean response for 500 randomly selected members of the 10,000 model runs. A.) Atmospheric methane concentration response for the default FUND version 3.9 (teal)  and modified FUND with methane feedbacks (purple). B.) Predicted discounted climate damage using a 3.0\% constant consumption discount rate.}
    \label{marginalresponses}
\end{figure}




\end{mytitlepage}


\begin{mytitlepage}
\baselineskip24pt
\setcounter{page}{1}
\title{{\Huge Supplementary Materials for} \\ \vspace{15pt}
Including climate system feedbacks in calculations of the social cost of methane} 

\author
{Kristina R. Colbert,$^{1\ast}$ Frank Errickson,$^{2}$ David Anthoff$^{2}$, \\ Chris E. Forest$^{1}$
\\
\normalsize{$^\ast$Correspondence to: cef13@psu.edu.}
}

\maketitle
\noindent \textbf{This PDF file includes:\\}
Methods\\
\indent Target Data Set for FaIR Calibration \\
\indent Updating the FaIR Natural Methane Emissions \\ 
\indent Updating the FaIR Methane Lifetime \\
\indent FaIR Calibration Process \\
\indent Uncertainty Estimation \\
\indent Social Cost of Methane Calculation \\
Supplementary Text \\
\indent Assessment of the Updated FaIR Methane Module \\
Supplementary Figures \\
\indent Captions for Figs. S1 to S5 \\
\indent Figs. S1 to S5 \\

\beginsupplement
\newpage
\section*{Methods}
We improve the FaIR methane module by introducing a natural wetland emissions component and new methane lifetime calculation. The new methane module successfully emulates the results of a higher complexity Earth system model, the MIT Earth System Model (MESM) \cite{Sokolov2018}, for a given emission scenario and reproduces observed methane concentrations.

We estimate the social cost of methane for a 1 t-\ce{CH4} emission pulse spread out over 10 years starting in 2020. The baseline emission trajectory follows the FUND built-in emission scenario. Methane concentrations change using our modified concentrations model which includes natural wetland emissions and feedbacks on the methane lifetime. Global and regional temperature, as well as, sea level changes lead to monetized impacts. Climate damages per region are summed through year 3000 using three different constant discount rate assumptions, 2.5\%, 3.0\%, and 5.0\%. The final SC-\ce{CH4} estimate sums all regional social cost estimates. We further align our SC-\ce{CH4} estimates with U.S. Government social cost of \ce{CO2} estimates by expressing the values in 2007 U.S. dollars per ton of methane.

\subsection*{Target Data Set for FaIR Calibration}
The methane module of FaIR, which calculates the change in methane concentration anomaly, is calibrated to a set of model simulations from the MIT Earth System Model (MESM) \cite{Sokolov2018}. The MESM links submodels of the atmosphere, ocean, terrestrial biosphere, and thermodynamic sea-ice to simulate critical feedbacks among its components, which are comparable to those of more complex Earth System Models \cite{Brasseur2016}. The MESM contains carbon cycle dynamics, biogeochemistry, and interactive atmospheric chemistry; critical components of an Earth System Model that provide an appropriate modeling system to assess methane dynamics. By including processes such as soil hydraulics, nutrient cycling, and carbon sequestration, the model includes an ecosystem component that is required to study the terrestrial carbon cycle. The other distinguishing component of the MESM model is its interactive atmospheric chemistry. Calculated on a daily timestamp, the atmospheric composition of climate-relevant gases and aerosols simulate the impact other trace atmospheric species have on methane's primary atmospheric sink \cite{Wang1998, Sokolov1998}. Climate system properties including climate sensitivity, rates of oceanic heat and carbon uptakes, and aerosol forcing are set to be consistent with available observations of surface air and ocean temperatures. We systematically vary the three parameters (climate sensitivity, rates of oceanic heat and carbon uptakes, and aerosol forcing) through a Latin Hypercube drawing of the joint probability distribution from \cite{Libardoni2019} to produce 49 parameter sets for an ensemble of MESM simulations.

Both natural and anthropogenic climate forcing agents drive the coupled system. In historical simulations, the model is first spun up by prescribed changes in greenhouse gases, aerosols, and solar irradiance from 1861 to 2005. Once the historical simulation is completed, the fully coupled-carbon chemistry model is initiated at year 2006 and driven by emission projections. To keep the emission scenarios between the MESM and FaIR consistent during the calibration process, we use the Emissions Prediction and Policy Analysis Model (EPPA) \cite{Paltsev2005} Outlook scenario \cite{Chen2016} as input into the FaIR model for the 2006 -- 2100 period. We also update FaIR to run with the new historical anthropogenic methane emissions used within the CMIP6, which is the Community Emissions Data System (CEDS) \cite{Hoesly2018}.

\subsection*{Updating the FaIR Natural Methane Emissions}

Run on an annual time-step, FaIR calculates the annual methane concentration (in parts per billion; ppb) as a balance between the sources and sinks.  Annual global emissions from anthropogenic ($A_t$) and natural sources ($N_t$) are converted into an equivalent increase in the atmospheric molar mixing ratio, $\delta C_t$ (1 ppb = 2.8403 Mt-\ce{CH4}). The concentration burden, $C_{t-1}$, is reduced by an exponential decay with a dynamic timescale dependent on the atmospheric concentration, $\tau_t$.

 \begin{equation}
C_t= C_{t-1} + {\frac{1}{2}(\delta C_{t-1}+\delta C_t )}- {C_{t-1} \left ( 1-\text{exp}\left (\frac{-1}{\tau_t}\right ) \right )}
\label{boxmodel1}
 \end{equation}

In the default FaIR setup, historic natural emissions are calculated to match observed concentrations by balancing the atmospheric budget to changes in anthropogenic emissions. The default model assumes future natural emissions are held at a constant 191 Mt-\ce{CH4} per year. As our first alteration to the FaIR model, we modify the natural emissions component to include a linear relationship between temperature and rate of wetland soil microbiological activity. The model calibration used the MESM output of global mean temperature output and global mean wetland methane emissions. The linear term for our updated global wetland methane emissions, $N_t$, is:

 \begin{equation}
N_t= m_N*(T_{t-1})+b_N
 \end{equation}
 
\noindent where $N_{t}$ is in Mt-\ce{CH4} yr$^{-1}$ and $T_{t-1}$ is the previous time-step's global mean temperature. $m_N$ is the slope and $b_N$ is the intercept of the linear function. The basis of this model came from preliminary analysis of MESM output for a 49 member ensemble (See Figure \ref{naturalvtemp}). The MESM results indicate a strong linear relationship (R$^2$=99\%) between increasing global mean surface temperature and increasing global wetland methane emissions. The relationship holds true for the historical period, as well as, projections for a business as usual future emission scenario. Methane emissions in the preindustrial period of the MESM span 172 to 198 Mt-\ce{CH4} yr\textsuperscript{-1}, falling within the uncertainty range of previous studies \cite{Paudel2016,Houweling2000}. MESM projections for end of the century global wetland emissions are 248--283 Mt-\ce{CH4} yr\textsuperscript{-1}, a substantial 43\% increase from preindustrial levels. Lacking the rise in future wetland emissions, the former FaIR v1.3 model underestimates a cumulative 3,973 Mt-\ce{CH4} of wetland emissions, on average, over the 21st century. Omitting such a large sum of wetland emissions is equivalent to omitting nearly 10 years of anthropogenic methane emissions (assuming 362--378  Mt-\ce{CH4} yr\textsuperscript{-1} from \cite{Saunois2016}). Figure \ref{49member_ch4nat} compares the emission's model fit to the MESM projections.

We acknowledge that the rate of natural emissions depends on several ecological and climate variables beyond just global mean temperature. For one, the amount of precipitation determines the water-saturation of the wetland soils, which is necessary to create anaerobic conditions for methanogenesis. With this in mind, we also examine the relationship between global precipitation and wetland methane emissions (Figure \ref{natemis_fit_precip}). As expected, the MESM results show global mean precipitation is strongly correlated with wetland methane emissions. While this shows precipitation may be a good predictor of natural methane emissions and could be used to build a simple model representation, precipitation is not a variable currently built into the FaIR model. To maintain the reduced-complexity structure of the FaIR model and to only focus on modifying the methane module, we do not model FaIR natural emissions as a function of precipitation. Our modification only models natural emissions to a change in simulated global mean temperature in FaIR.

\subsection*{Updating the FaIR Methane Lifetime}
We next alter the FaIR model sink term by modifying the defaulted constant 9.3 year methane lifetime so that the methane lifetime parameter takes into account changes in methane abundance that lead to fluctuations in its primary sink, OH. We calibrate the methane lifetime parameters to reproduce the observed changes in methane concentration over the historical period and future projections produced by the MESM. In the MESM, methane lifetime is not a prescribed parameter, but instead, is a property that emerges from the model chemistry \cite{Crutzen1991}. This allows the MESM to simulate the complex chemical reaction sequence following methane oxidation, as well as, changes in the methane lifetime feedback to OH concentrations. For our purposes, the MESM only considers the tropospheric OH sink; it does not model the methane sink to chlorine, soils, and stratosphere.

Our modification aims to maintain the box model representation in FaIR to preserve the reduce-complexity structure. Our approach solves for the methane lifetime in the FaIR mass balance equation by providing all of the other known quantities of the equation with MESM data. That is to say, we use the MESM concentrations data ($C_{t}$), and emissions data (converted to molar mixing ratios; $\delta C_{t}$) to solve for the methane lifetime, $\tau$, at each time-step. In essence, this recalculates the methane lifetime required to achieve the observed change in methane concentration over the historical period and future projections produced by the MESM. We do this for each of the 49 MESM ensemble members. 

Figure \ref{lifetime_budget_MESM} shows the resulting time-series for the methane lifetime. The jagged shape reflects major changes in the historic methane budget, as the lifetime responds to changes in emissions. Our preindustrial estimates for methane's lifetime range higher than the original FaIR model assumption, averaging 9.8 $\pm$ 0.2 years before the late 1970s. The 1980s and early 1990s are marked by large fluctuations and peaks in methane's lifetime following volcanic events such as the eruptions of Mt. St. Helens (1980), El Chich\'{o}n (1982), and Pinatubo (1991) \cite{John2012}. The 1980s to 2000s show a decreasing trend in methane lifetime in the MESM, agreeing with the findings of \cite{Arora2018}. As a final check, our present-day (for the year 2000) methane lifetime estimate is 9.5 $\pm$ 0.1 years, well within the range of ACCMIP models of 9.7 $\pm$ 1.5 years \cite{Naik2013} and recent observation-based calculations for 2010, 9.1 $\pm$ 0.9 years \cite{Prather2012}. Using the MESM future emissions projections, we can also predict how methane lifetime may evolve over time. The future projections indicate a steep incline in methane lifetime, increasing from just below 9.0 years to nearly 12.8 years by the end of the century. This demonstrates the deficiencies in a constant lifetime assumption and highlights the utility of a chemistry model for calibration of a reduced-complexity model.

The complex nature of methane's calculated lifetime calls for a piecewise-defined function approach, where the methane lifetime, $\tau$, over time, $t$ is:

\begin{equation}
\tau(t) = \begin{cases} 
          9.8 & t\leq 1981 \\
          \beta_0 + \beta_1 * t & 1981\leq t\leq 2008 \\
          exp [k_0 + k_1 * log(C_{t-1})] & 2008\leq t 
       \end{cases}
       \label{lifetime_piecewise1}
\end{equation}

From the model initiation until 1981, a constant lifetime is assumed to be 9.8 yr. The second subdomain, which corresponds to the plateau in methane concentrations up until 2008, uses a linear time series regression of the predicted year, $t$. These two assumptions accurately reproduce historic observed global mean methane concentrations. For present-day and future projections of methane emissions, the lifetime is a function of atmospheric methane concentration. Thus, the new lifetime term improves the model by incorporating feedback effects on the methane sink with increasing concentrations.

\subsection*{FaIR Calibration Process}
The purpose of the calibration is to find a set of parameter values, $\theta$, that optimizes the fit of the FaIR simulated data to an "observation" data set, or target data set, from the MESM. The MESM time-series output used to constrain FaIR includes annual global means of (1) atmospheric methane concentration, (2) methane lifetime, and (3) natural methane emissions from wetlands over the 1861-2100 period. For the calibration purpose, the methane module is isolated from the remainder of the FaIR components. 

To span the uncertainty caused by several different climate states, we calibrate the FaIR methane parameters to each of the 49 ensemble members from the MESM using a Markov Chain Monte Carlo (MCMC) method \cite{Metropolis1953,Hastings1970}. While there are several variants of the MCMC, we use the Robust Adaptive Metroplis (RAM) algorithm \cite{Vihola2012}. The RAM allows for a more efficient way to estimate the target distribution by updating the proposal distribution with each sampling. We also account for potential residual autocorrelation in the model error and time-varying heteroskedastic observation errors, by estimating the residuals with a stationary first-order autoregressive process AR(1) model \cite{Ruckert2017}.

The initial guess for the the six new uncertain parameters is based on some previous knowledge, in this case, the preliminary regression fits for natural wetland emissions and methane lifetime. We assume a uniform probability density for the prior distributions. We use 100,000 iterations and remove 10\% of the initial "burn-in" from the starting values of the Markov chain. The burn-in is to ensure that the results are not dependent on the initial conditions. In addition, we "thin" the sample chains to a 1,000 iteration segment to produce more independent values for the analysis. Combining the thinned Markov chain results for each of the 49 calibrations, we arrive at a total of 49,000 independent parameter set combinations.

\subsection*{Uncertainty Estimation}
Producing sound estimates of SC-\ce{CH4} requires spanning the full range of the FUND parameter space. We conduct a formal probabilistic analysis through random-sampling Monte Carlo simulations \cite{Metropolis1953,Hastings1970}. The Monte Carlo sampling technique is a classic procedure of repeating random draws from a defined probability distribution of a variable. A sufficient number of draws will result in a posterior distribution that reflects the shape of the prior defined distribution of the random variable.

We perform a Monte Carlo sampling of 10,000 samples on each of the 852 uncertain parameters of the modified FUND model. To include the uncertainty in the selection of the discount rate, we repeat the Monte Carlo sampling for simulations using a constant discounting assumption of 2.5\%, 3.0\%, and 5.0\%. With 10,000 posterior parameter samples per discount rate yields 30,000 unique SC-\ce{CH4} estimates. 


\subsection*{Social Cost of Methane Calculation}
The social cost estimate is the monetary measure of the aggregated damage done by a unit-ton greenhouse gas emission in a given year \cite{IWG2017}. Two climate trajectories are used for the estimation of the SC-\ce{CH4}; a baseline trajectory that follows a set emission scenario, and a pulse emission trajectory that demonstrates the perturbation response to one ton of global methane emissions, spread out over 10 years in FUND \cite{Waldhoff2014}. The response in global temperature and sea level result in monetary damages. The difference between the two damage trajectories is the annual marginal climate damages, $D_{t,r}$ (in 2007 U.S. dollars per year), calculated at each region, $r$, per year, $t$:

\begin{equation}
    D_{t,r} = D_{{pulse}_{ t,r}} - D_{{baseline}_{ t,r}}
\end{equation}

The marginal damages are discounted back to present-day values using a chosen discount rate. The discount rate weights how much future costs or benefits are considered to be less significant than present costs and benefits. This is analogous to the idea if you become richer over time, extra costs or benefits have less value or impact to you than when you were poorer. 

The discount rate, $r$, is inherently related to economic growth, $g$, as demonstrated by the Ramsey formula \cite{Ramsey1928},

\begin{equation}
    r = \delta + \eta*g
\end{equation}

\noindent where $\delta$ is the pure time preference rate, $\eta$ is the elasticity of marginal utility of consumption, and $g$ is the economic growth per capita. As discussed in \cite{IWG2017}, $\delta$ reduces future values 0--3\% per year, $\eta$ measures aversion to differences in consumption between individuals today or generations across time, and $g$  reflects the monetary value of all market goods and services over the past few decades. 

To compare SC-\ce{CH4} estimates cited by the U.S. Government \cite{IWG2017}, all of our results assume a constant consumption discount rate, by holding $\eta$ at zero. Because the discount rate compounds impacts over time, small differences in the selected discount rate can have large impacts on the social cost estimate. For this reason, we explore the SC-\ce{CH4} estimated using three suggested discount rates used in the estimation of the SCC, 2.5\%, 3.0\%, and 5.0\% \cite{Marten2015}.

In FUND, the social cost is first estimated per region, $SC_r$ by solving for the discounted sum of the annual marginal damages, $D_{t,r}$, normalized by the pulse emission, $\epsilon$, for all time following the pulse emission:

\begin{equation}
    SC_r =  \dfrac{1}{\epsilon} \sum_{t=2020}^{3000}  \dfrac{D_{t,r}}{(1+ r)^t}
\end{equation}

The final social cost estimate, $SC$, is the aggregate of the 16 regional social cost estimates, $SC_r$. 

\begin{equation}
    SC =  \sum_{r=1}^{16} SC_r
\end{equation}

To review, we estimate the social cost of methane for a 1 t-\ce{CH4} emission pulse spread out over 10 years starting in 2020. The baseline emission trajectory follows the built-in emission scenario, IS92a. Methane concentrations change using our modified concentrations model which includes natural wetland emissions and feedbacks on the methane lifetime. Global and regional temperature, as well as, sea level changes lead to monetized impacts. Climate damages per region are summed through year 3000 using three different constant discount rate assumptions, 2.5\%, 3.0\%, and 5.0\%. The final SC-\ce{CH4} estimate sums all regional social cost estimates. We further align our SC-\ce{CH4} estimates with U.S. Government social cost of \ce{CO2} estimates by expressing the values in 2007 U.S. dollars per ton of methane.

\section*{Supplementary Text}
\subsection*{Assessment of the Updated FaIR Methane Module}
We run the default FaIR v1.3 model, as well as, the updated methane module with a new natural emissions model (with the defaulted constant lifetime), the updated module with a new lifetime calculation (but no new natural emissions), and the fully updated model, containing the natural emissions model and lifetime calculation. Each of the FaIR model versions is run using the same initial conditions, emissions inputs and maximum likelihood estimates for the uncertain parameters. In Figure \ref{49conc}, the updates for each component of methane model are compared against the MESM ensemble and observations. Derived observational data sets of global atmospheric methane concentrations from Mauna Loa and Law Dome ice cores are plotted as a reference for the historical period.

Exploring one update at a time allows for the isolation of the independent impacts and combined impacts of including a new natural emissions model and concentration-driven lifetime. First looking at the default FaIR v1.3 (shown in red in Figure \ref{49conc}), the default model poorly matches the historical observations. This underestimated concentration from the default model is due to the fact that its concentrations were originally calibrated to match a higher assumed historic anthropogenic methane emissions (i.e. historic RCP scenario) than that used by the CEDS historical emissions scenario. But what is notable is the default model greatly underestimates the projected future methane concentration, falling 903 ppb below the mean projection of the MESM ensemble in 2100. Clearly, the default model for the methane cycle requires an update to better match projections of a higher complexity Earth System Model.

The first update explored was replacing the prescribed external natural methane emissions data set in the default model with a natural wetland emissions model dependent on the global temperature. Replacing only the natural emissions component of the default model improves the concentration projections significantly for the historical period (Figure \ref{49conc}). The higher natural methane emissions calculated in the new model raises the total source term of the box model calculation, allowing for higher methane concentrations. This change alone does not sufficiently match future methane projections, a change to the lifetime calculation is the next logical step. However, replacing the constant lifetime assumption in the default model with a dynamic methane lifetime dependent on atmospheric burden, alone, also does not fully capture the methane projections. The profile and shape of the curve (green curve in Figure \ref{49conc}) resemble that of the MESM, but the magnitude of the concentration still falls below the target data set for the entirety of the simulation. 

Combining both the new natural emissions model and lifetime calculation into the FaIR methane module provides the best representation of the MESM target data set. Visual inspection indicates that the updated FaIR model accurately captures the historical observations and future projection of methane from the MESM. The ensemble of FaIR closely matches the ensemble uncertainty of the MESM. There are only two periods in which the updated model shows a deviation from the target data set; in the early 1900s and 2000s where the deviation reaches up to 25 ppb. This deviation is primarily due to the FaIR lifetime not matching the simulated MESM lifetime. Checking the Pearson correlation coefficient ($r^2$ = 0.999) as a statistical standard, we concluded that the updated FaIR model can accurately predict the methane concentration from the prescribed anthropogenic emissions scenario.


\clearpage
\section*{Supplementary Figures}
\noindent {\bf Fig. S1.} Mean annual methane emissions from wetlands within the MESM model as a linear function of global mean surface temperature anomaly. The colored scale refers to the modeled year.

\noindent {\bf Fig. S2.} Comparisons of global wetland methane emissions. Shown are the default FaIR model (red), the MESM ensemble target data set (gray) and the updated FaIR model using a temperature-driven natural emissions model (blue). Displayed are the full ranges for the ensembles and their mean ensemble estimates.

\noindent {\bf Fig. S3.} Mean annual methane emissions from wetlands within the MESM model as a linear function of global mean precipitation. The color scale indicates the modeled year.

\noindent {\bf Fig. S4.} A time series of the calculated methane lifetime for the default FaIR model (red), the 49 member ensemble of the MESM (gray), and an example of a fitted model for our modified FaIR lifetime (blue).

\noindent {\bf Fig. S5.} Global mean methane concentration simulated by the FaIR default v1.3 model (red) and the methane model updates. Model updates include the natural emissions model alone (orange), concentration-driven lifetime model alone (green), and the two combined to give the fully updated model (light blue). The MESM 49 member ensemble is shown with its mean response (gray). And observations from Mauna Loa and Law Dome are shown (purple).


\newpage
\begin{figure}[b!]
\centering
\includegraphics[width=\linewidth]{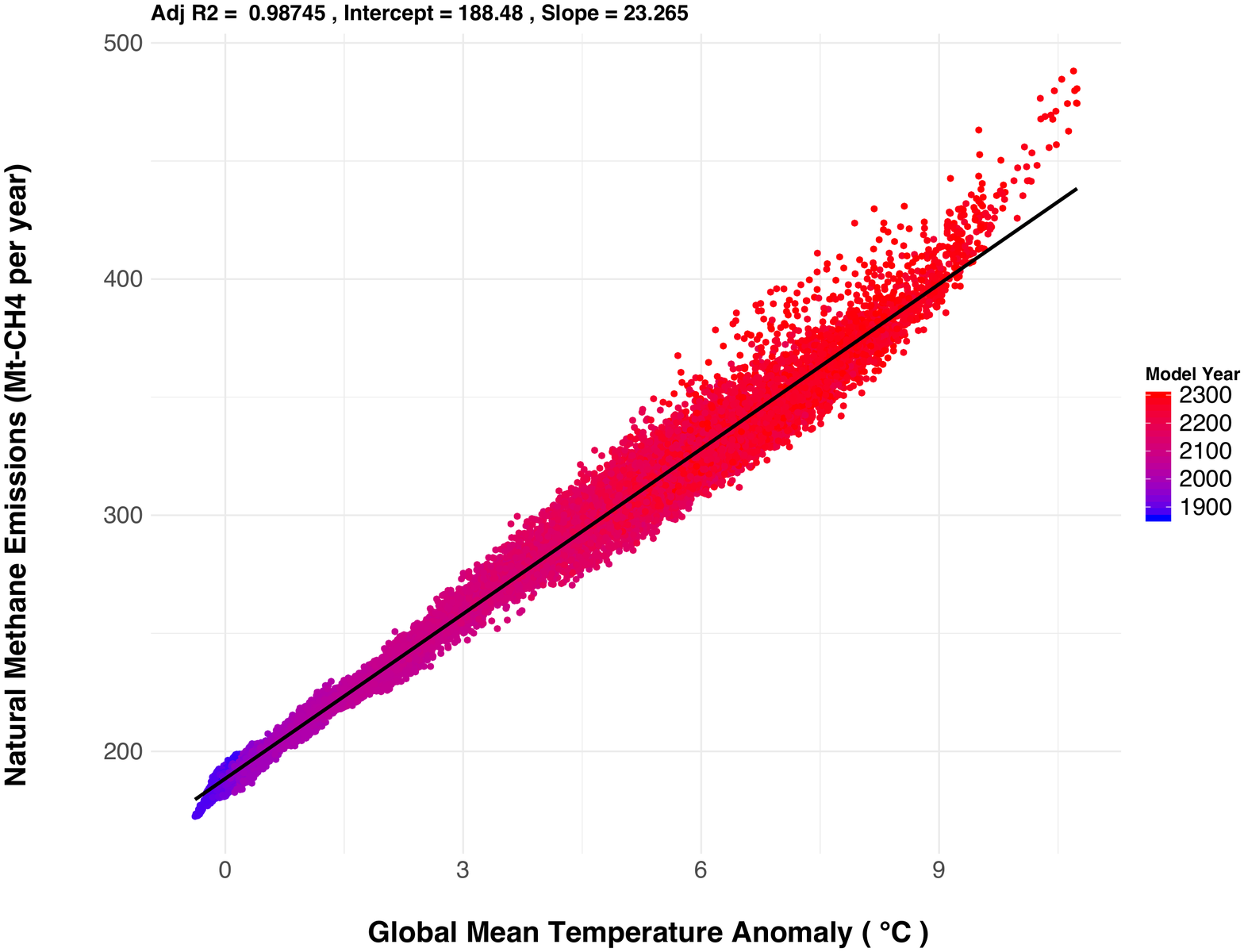}
\caption{Mean annual methane emissions from wetlands within the MESM model as a linear function of global mean surface temperature anomaly. The colored scale refers to the modeled year.} 
\label{naturalvtemp}
\end{figure}

\newpage
\begin{figure}[hb!]
\centering
\includegraphics[width=\linewidth]{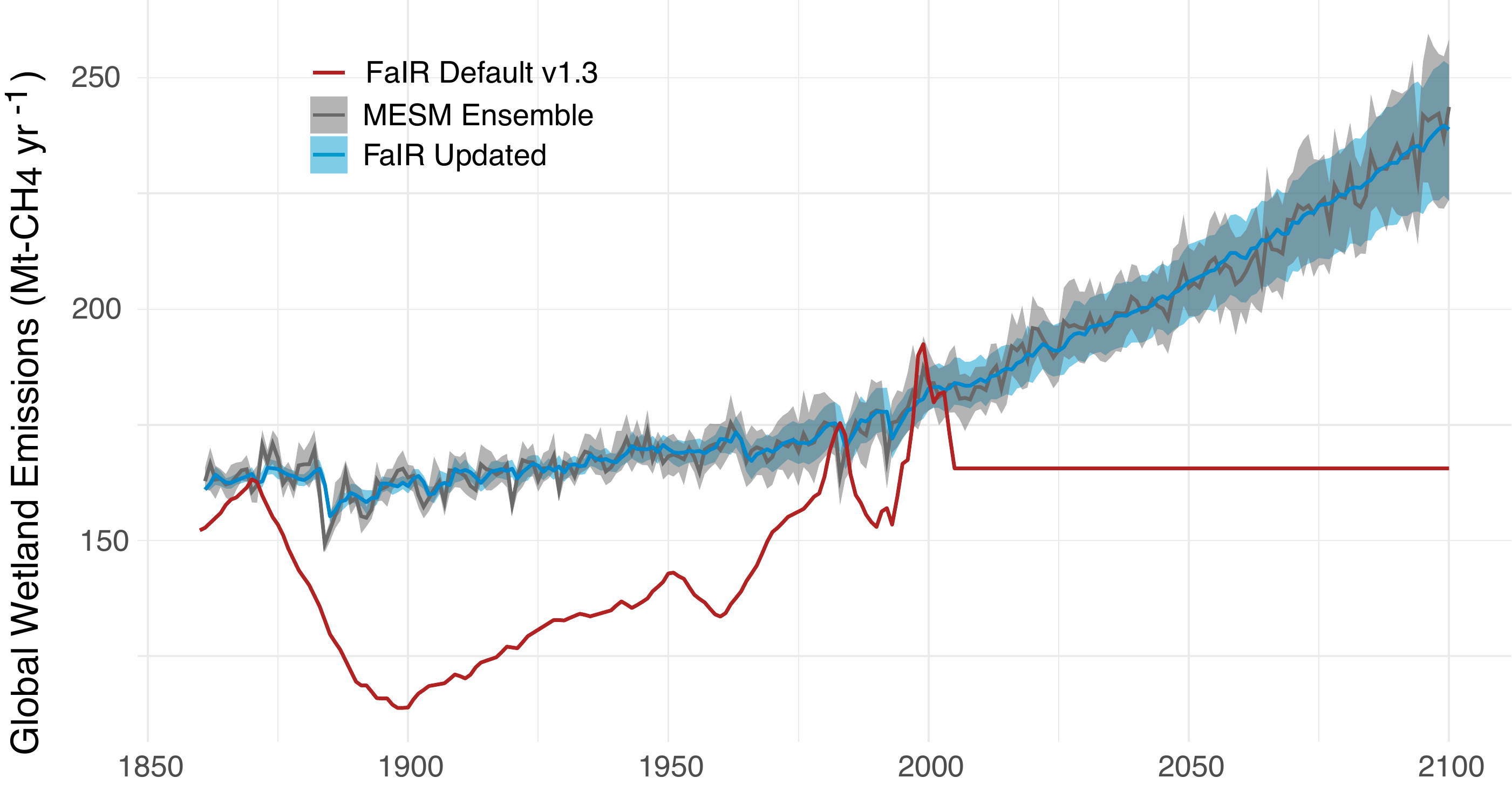}
\caption{Comparisons of global wetland methane emissions. Shown are the default FaIR model (red), the MESM ensemble target data set (gray) and the updated FaIR model using a temperature-driven natural emissions model (blue). Displayed are the full ranges for the ensembles and their mean ensemble estimates.}
\label{49member_ch4nat}
\end{figure} 

\newpage
\begin{figure}[hb!]
\centering
\includegraphics[width=\linewidth]{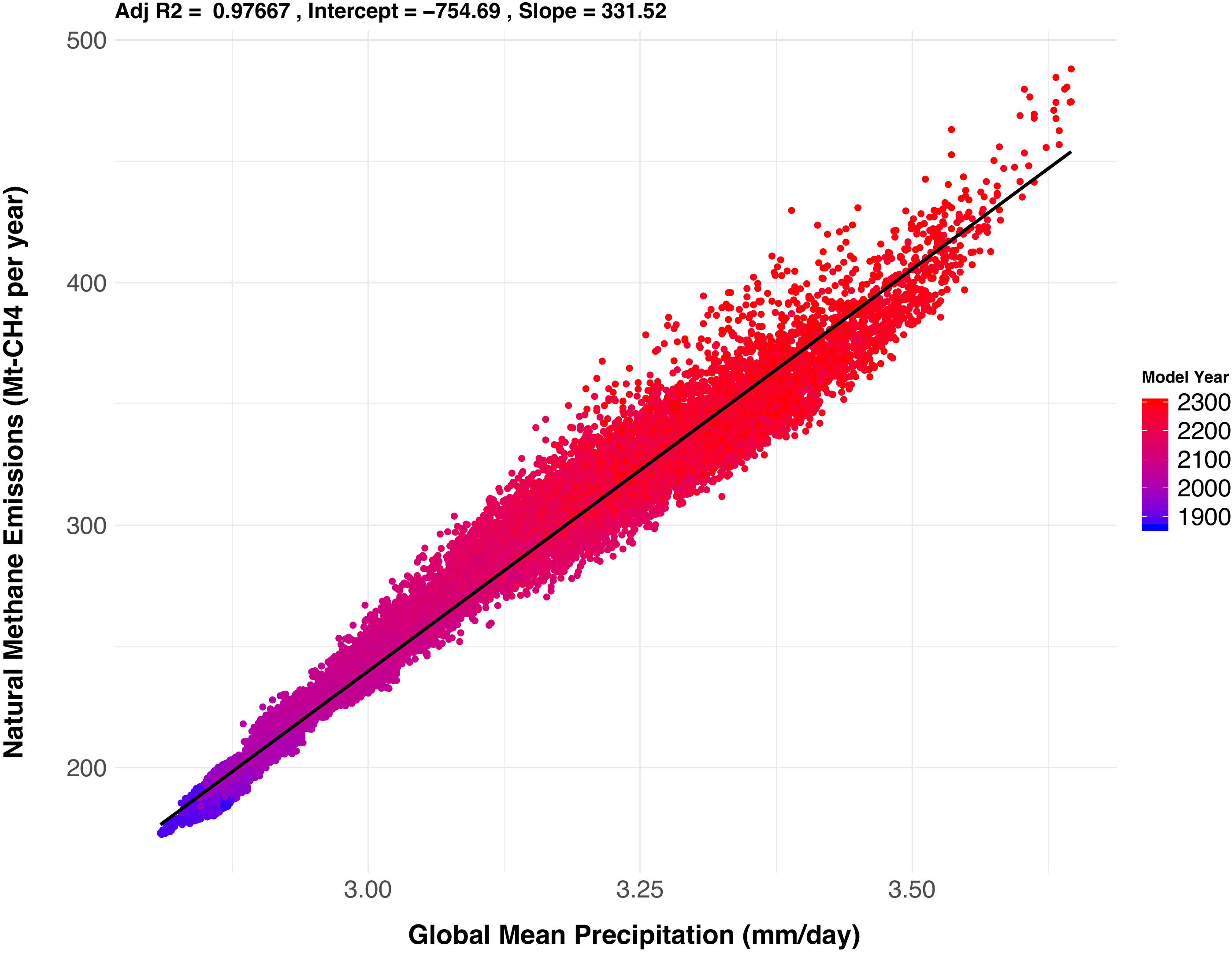}
\caption{Mean annual methane emissions from wetlands within the MESM model as a linear function of global mean precipitation. The color scale indicates the modeled year.}
\label{natemis_fit_precip}
\end{figure} 

\newpage
\begin{figure}[hb!]
\centering
\includegraphics[width=\linewidth]{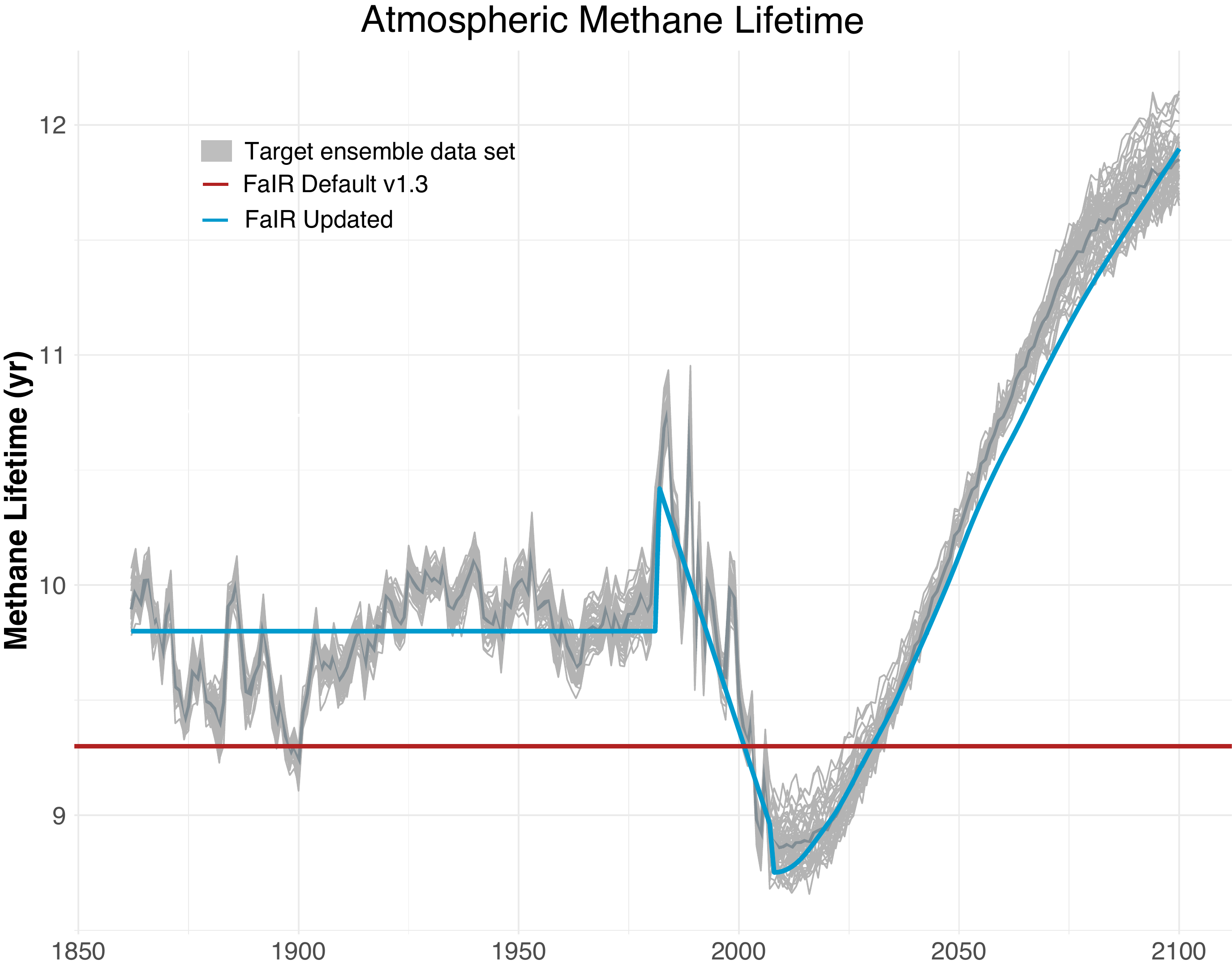}
\caption{A time series of the calculated methane lifetime for the default FaIR model (red), the 49 member ensemble of the MESM (gray), and an example of a fitted model for our modified FaIR lifetime (blue).}
\label{lifetime_budget_MESM}
\end{figure}

\newpage
\begin{figure}[hb!]
\centering
\includegraphics[width=\linewidth]{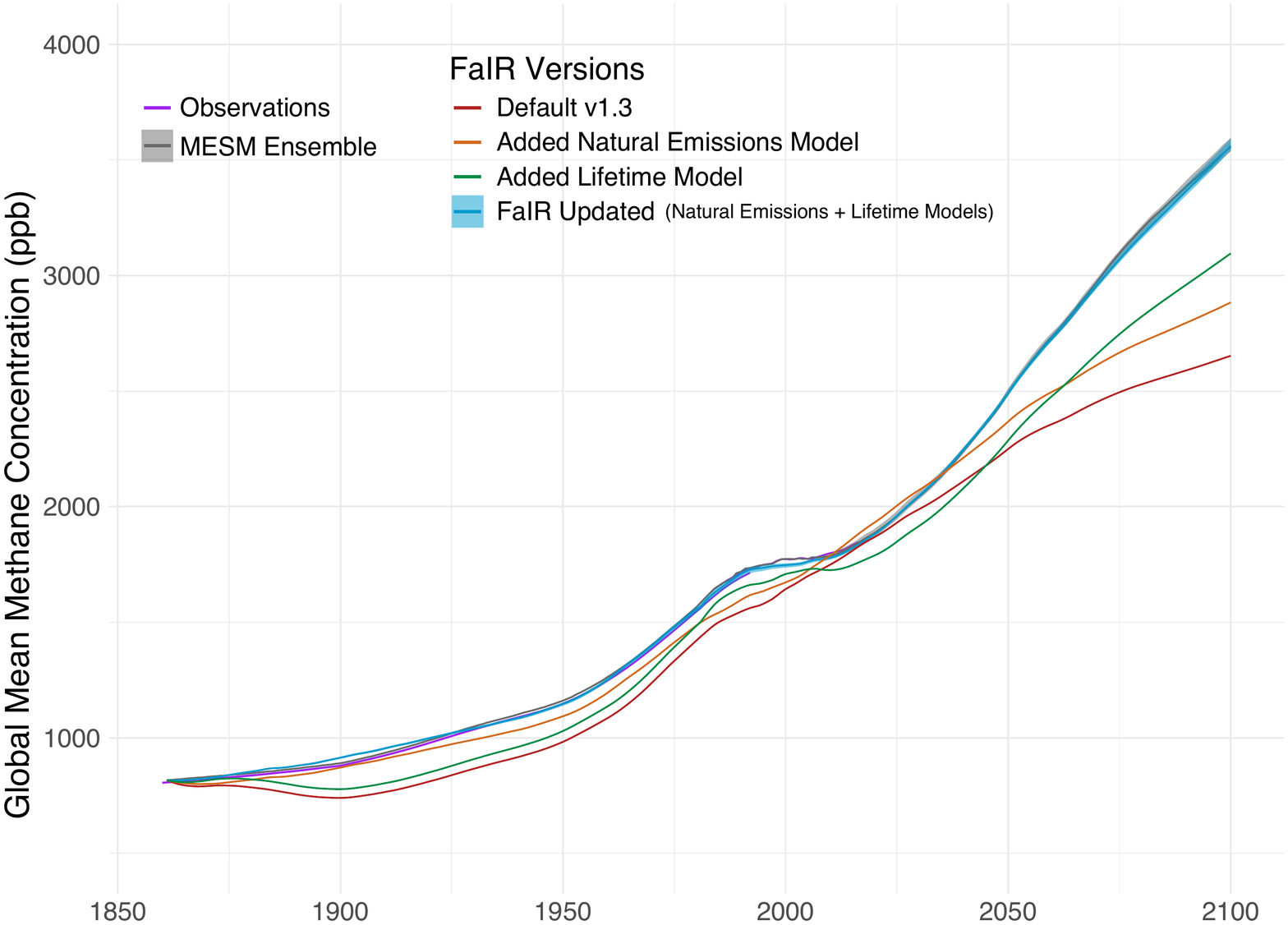}
\caption{Global mean methane concentration simulated by the FaIR default v1.3 model (red) and the methane model updates. Model updates include the natural emissions model alone (orange), concentration-driven lifetime model alone (green), and the two combined to give the fully updated model (light blue). The MESM 49 member ensemble is shown with its mean response (gray). And observations from Mauna Loa and Law Dome are shown (purple).} 
\label{49conc}
\end{figure}


\end{mytitlepage}

\end{document}